\documentclass[%
 reprint,
 superscriptaddress,
 amsmath,amssymb,
 aps,
]{revtex4-1}

\usepackage{comment, color}
\usepackage{braket}

\usepackage{graphicx}% Include figure files
\usepackage{dcolumn}% Align table columns on decimal point
\usepackage{bm}% bold math
\begin{document}

\title{Second-order topological phases protected by chiral symmetry}% Force line breaks with \\

\author{Ryo Okugawa}
 \affiliation{%
 	WPI-Advanced Institute for Materials Research (WPI-AIMR), Tohoku University, 2-1-1, Katahira, Sendai 980-8577, Japan
 }%
\author{Shin Hayashi}
\affiliation{%
	Mathematics for Advanced Materials-OIL, AIST, 2-1-1
	Katahira, Aoba, 980-8577 Sendai, Japan
}%
\affiliation{%
	JST, PRESTO, 4-1-8 Honcho, Kawaguchi, Saitama, 332-0012, Japan
}%
\author{Takeshi Nakanishi}
\affiliation{%
	Mathematics for Advanced Materials-OIL, AIST, 2-1-1
	Katahira, Aoba, 980-8577 Sendai, Japan
}%

\date{\today}% It is always \today, today,

\begin{abstract}
We study second-order topological insulators and semimetals characterized by chiral symmetry.
%We study second-order topological semimetals and insulators characterized by chiral symmetry.
We investigate topological phase transitions of a model
for construction of the two-dimensional second-order topological insulators protected only by chiral symmetry.
By the theory of the phase transitions,
we propose a second-order topological semimetal and insulators with flat hinge bands in chiral-symmetric three-dimensional systems.
%in view of the topological phase transitions for two-dimensional second-order topological insulators protected by chiral symmetry.
The three-dimensional second-order topological phases can be obtained
from the stacked two-dimensional second-order topological insulators with chiral symmetry.
Moreover, we show that broken chiral symmetry in the three-dimensional second-order topological phase
allows a second-order topological insulator with chiral hinge states. 
We also demonstrate the second-order topological phases by using a lattice model.
\end{abstract}

\maketitle

\section{Introduction}
Higher-order topological insulators have recently drawn research interest as new topological crystalline phases
\cite{Benalcazar17S, Benalcazar17B, Song17, Fang17, Schindler18, Langbehn17, Geier18, Luka19, Khalaf18X, Khalaf18B, Matsugatani18, Miert18, Kooi18, Wang18, Ezawa18B, Ezawa18L, Ezawa18L2, Ezawa18B2, Ezawa18B3, Ezawa19, Fukui18, Kunst18, Franca18, Dumitru19,
Schindler18N, Imhof18, Serra18, Peterson18, Ni19, Fan19, Ahn19, Liu17, Liu19, Rodriguez19, Benalcazar19, Wieder18, Wieder18s, Hwang19}.
Unlike conventional first-order topological insulators,
two-dimensional (2D) second-order topological insulators (SOTIs) have topologically protected corner states,
and three-dimensional (3D) SOTIs have topological gapless modes on the hinges.
%The nontrivial topology in many SOTIs stems from the crystal symmetry.
In some crystalline SOTIs, the topological corner and hinge states can arise only
from the nontrivial bulk topology,
%characterized by the crystal symmetries
when the lattice termination is compatible with the crystal symmetries.
The crystalline insulators with the nontrivial second-order topology, which are independent of the lattice termination,
are called intrinsic second-order topological phases \cite{Langbehn17, Geier18, Luka19}.

Intriguingly, stacking 2D SOTIs enables the coexistence of topological gapless surface and hinge states
while the bulk gap is open.
The 3D SOTI can be realized by two anticommuting mirror symmetries \cite{Lin18}.
Furthermore, the idea of the crystalline SOTIs has been extended to semimetallic phases,
which are called second-order topological semimetals (SOTSMs)
\cite{Lin18, Wang18, Ahn18, Ezawa18B, Ezawa18L, Ezawa19, Dumitru19, Wieder19}.
SOTSMs have not only topological hinge states
but also topological gapless nodes in the bulk.
The topological hinge states appear between the gapless points projected onto the hinges.

On the other hand, crystal symmetry is not necessarily required in order to realize SOTIs
\cite{Seradjeh08, Volovik10, Sitte12, Zhang13, Hashimoto17, Li18, Langbehn17, Geier18, Luka19, Chen19, Bomantara19, Hayashi18, Hayashi19}.
In this case,
the topological classification of the $d$-dimensional second-order topological phases is the same
as that of the $(d-1)$-dimensional first-order topological phases
in the Altland-Zirnbauer classes \cite{Langbehn17, Geier18, Luka19, Hayashi18, Hayashi19}.
The classification is based on the $K$ theory \cite{Langbehn17, Geier18, Luka19, Hayashi18, Hayashi19},
which means that the nontrivial phases are not fragile.
Physically, the corner and hinge states are understandable as domain wall modes between topologically trivial and nontrivial boundaries
\cite{Langbehn17, Geier18, Luka19, Teo10}.
Thus, the SOTIs in the 3D class A and the 2D class AIII are characterized by $\mathbb{Z}$ invariants,
which depend on the crystal termination.
Such nontrivial phases are named extrinsic SOTIs \cite{Langbehn17, Geier18, Luka19},
whose topology of the corner (hinge) states is determined by both the bulk and the edges (surfaces).
%In the 2D (3D) SOTIs, both the bulk and the edges (surfaces) determine topology of the corner (hinge) states.
The SOTI phases are realizable 
even if we arbitrarily terminate the lattice \cite{Langbehn17, Geier18, Luka19,  Hayashi18, Hayashi19, Takane19}.
The topological modes are stable as long as the bulk and the boundary bands are gapped.

However, we need to calculate a complicated boundary Hamiltonian to study extrinsic second-order topology in general.
The phase diagram to search for the SOTI is difficult
to explore analytically without employing crystal symmetry.
Thus, it is useful to give a systematic way to construct extrinsic SOTIs and the explicit topological invariant for the phase diagram.
Moreover, topological gapless nodes are characterized by a change of topological invariants for lower-dimensional insulators
in momentum space.
Therefore, we can produce new 3D second-order topological phases from the phase diagram for the 2D SOTIs.
%Because SOTIs with in crystal  have not been revealed yet.

In this work, we show various second-order topological phases realized by chiral symmetry.
We discuss a simple method to construct the chiral-symmetric second-order topological phases.
To do so, we develop the previous method \cite{Hayashi18, Hayashi19} for the 2D SOTI protected only by chiral symmetry
in view of the topological phase transition.
From our theory, we also show different 3D second-order topological phases due to chiral symmetry,
and one of them is an unconventional topological phase which has a single gapless point on the 2D surface
in addition to the topological hinge states.
Furthermore, it is found that the 3D SOTI with surface gapless points can become
a 3D SOTI with chiral hinge states by broken chiral symmetry.

This paper is organized as follows.
In Sec.~\ref{SOTP},
we show the method for the construction of the second-order topological phases due to chiral symmetry.
We demonstrate the topological phases by a lattice model constructed from our method in Sec.~\ref{model}.
Our conclusion is summarized in Sec.~\ref{con}.

\section{Second-order topological phases characterized by chiral symmetry} \label{SOTP}
In this section, we study SOTIs and a SOTSM characterized by chiral symmetry
in terms of the bulk and edge gap closings for the topological phase transitions.
We give a simple Hamiltonian to create the 2D SOTIs \textit{only by chiral symmetry}
and extend it to 3D systems by considering the translation symmetry.

\subsection{Corner states and topological invariant}
First of all, we review construction of 2D SOTIs in class AIII according to Ref.~\onlinecite{Hayashi18}.
We apply the method to 3D systems with chiral symmetry later.
We consider the following bulk Hamiltonian:
\begin{align}
	H(\bm{k})=\mathcal{H}_x(k_x)\otimes \Pi _y+ 1_x \otimes \mathcal{H}_y(k_y), \label{kH}
\end{align}
where $\mathcal{H}_{i}(k_{i})~(i=x,y)$ are two Hermitian matrices with chiral symmetry represented as $\Pi _{i}$.
Here, $\Pi _{i}^2=1_i$, and $1_i$ is the identity matrix with the same size as $\mathcal{H}_{i}(k_i)$.
Because $\{ \mathcal{H}_i, \Pi _i\} =0$ is satisfied,
the Hamiltonian in Eq.~(\ref{kH}) has chiral symmetry $\Pi = \Pi _x\otimes \Pi _y$.
In this paper, we focus on gap closing at zero energy
because chiral symmetry is present.
We impose open boundary condition (OBC) with a right-angled corner \cite{comment}.
When the model is gapped at zero energy,
it can be characterized by a $\mathbb{Z}$ topological invariant given by \cite{Hayashi18, Hayashi19}
\begin{align}
	\nu _{2D}=w_x w_y, \label{inv} 
	\end{align}
where $w_{i=x,y}$ are conventional winding numbers for $\mathcal{H}_{i}(k_i)$. 
The expression of the winding number is given in the Appendix \ref{winding}.
The winding numbers distinguish
whether the one-dimensional (1D) bulk Hamiltonians $\mathcal{H}_{i}(k_i)$ are
in the first-order topological phase by chiral symmetry \cite{Ryu02, Schnyder08, Ryu10, Chiu16}.
As discussed in the next section,
$\nu _{2D}$ is unchanged as long as the model is gapped at zero energy in the bulk and on the edges.
The system with the nonzero $\nu _{2D}$ is a 2D SOTI with zero-energy corner states. %\cite{comment1}.
%(Instead of Eq.~(\ref{kH}), we can use $H(\bm{k})=H_x(k_x) \otimes 1_y + \Pi _x \otimes H_y(k_y)$.)

Indeed, the nonzero $\nu _{2D}$ indicates the existence of the corner states
when the edges break the translation symmetries in the $x$ and the $y$ directions.
We consider a semi-infinite system with one corner for Eq.~(\ref{kH}).
Because $k_x$ and $k_y$ are separate in Eq.~(\ref{kH}),
the real-space Hamiltonian can be described
by two real-space Hamiltonians of $\mathcal{H}_x(k_x)$ and $\mathcal{H}_y(k_y)$.
We denote the Hamiltonians of $\mathcal{H}_i(k_i)$ as $\mathcal{H}_i^{\mathrm{OBC}}$.
The Hamiltonian under the corner boundary condition (CBC) can be represented as \cite{comment}
\begin{align}
	{H}^{\mathrm{CBC}}=\mathcal{H}_x^{\mathrm{OBC}} \otimes \Pi _y^{\mathrm{OBC}}
	+ 1_x^{\mathrm{OBC}} \otimes \mathcal{H}_y^{\mathrm{OBC}}. \label{cornerH}
\end{align}
Here, $\Pi _y^{\mathrm{OBC}}$ and $1_x^{\mathrm{OBC}}$ are representations of the chiral symmetry and the identity matrix
in the terminated system, respectively.
To see topological corner states, we assume that the bulk and the edges are gapped at zero energy.
If both $\mathcal{H}_x(k_x)$ and $\mathcal{H}_y(k_y)$ are topologically nontrivial, i.e. $\nu _{2D}\neq 0$,
we can see the corner states with the zero energy as follows.
Let $\phi _i^{zero}$ be one of the topological zero-energy eigenvectors of $\mathcal{H}_i^{\mathrm{OBC}}$
at one zero-dimensional edge from the nonzero $w_i$.
Then, we can find a zero-energy state given by $\phi _x^{zero} \otimes \phi _y^{zero}$ for ${H}^{\mathrm{CBC}}$ in Eq.~(\ref{cornerH}).
By assumption, the zero-energy state should be a corner state.
Generally, we can obtain $|\nu _{2D}|=|w_xw_y|$ topological corner states
because $\mathcal{H}_{i=x,y}^{\mathrm{OBC}}$ have $|w_i|$ zero-energy modes. 
Thus, zero-energy corner states appear when $\nu _{2D}$ is nonzero.

\subsection{Topological phase transitions and gap closing}
To grasp the 2D SOTI,
we revisit the topological phase transitions in view of the gap closing.
We clarify how the topological phases for $H(\bm{k})$ change when we continuously deform $\mathcal{H}_i(k_i)$.
In other words, we can see how the band gap necessarily closes in the bulk or the edges when $\nu _{2D}$ changes.

We assume that the system is in a trivial phase with $\nu _{2D} = 0$ to elucidate the topological phase transitions.
$\mathcal{H}_x(k_x)$ and $\mathcal{H}_y(k_y)$ need to close the band gap to change $\nu _{2D}=w_xw_y$.
To begin with, we discuss the bulk gap given by Eq.~(\ref{kH}).
Because $\{ \mathcal{H}_y, \Pi _y\} =0$, the bulk Hamiltonian satisfies
\begin{align}
	H(\bm{k})^2= \mathcal{H}_x(k_x)^2 \otimes 1_y + 1_x \otimes \mathcal{H}_y(k_y)^2. \label{HH}
	\end{align}
Therefore, if and only if $\mathcal{H}_x(k_x)$ and $\mathcal{H}_y(k_y)$ take zero-valued eigenvalues at the same time, 
the bulk bands close the gap.
This gap closing can change both $w_x$ and $w_y$, and thus $\nu _{2D}$.

By contrast, $\nu _{2D}$ can change
even if $\mathcal{H}_x(k_x)$ and $\mathcal{H}_y(k_y)$ do not take zero-valued eigenvalues simultaneously.
We consider a trivial phase with $(w_x, w_y)=(1,0)$.
From the above discussion, 
when the system enters a nontrivial phase with $(w_x, w_y)=(1,1)$ from the trivial phase,
the bulk Hamiltonian $H(\bm{k})$ does not close the band gap.
Then, we investigate an edge normal to the $x$ direction.
Namely, we see a semi-infinite system with the edge along the $y$ direction.
Because we retain periodicity in the $y$ direction,
the Hamiltonian under the edge boundary condition (EBC) can be described as
\begin{align}
	{H}^{\mathrm{EBC}}_x(k_y)=\mathcal{H}_x^{\mathrm{OBC}} \otimes \Pi _y
	+ 1_x^{\mathrm{OBC}} \otimes \mathcal{H}_y(k_y). \label{edgeHy}
	\end{align}
%which satisfies
%\begin{align}
%	[{H}^{edge}_x(k_y)]^2=[\mathcal{H}_x^{edge}]^2 \otimes  1_y+ 1_x^{edge} \otimes {H}_y(k_y)^2.
%	\end{align}
Although ${H}^{\mathrm{EBC}}_x(k_y)$ satisfies a condition similar to Eq.~(\ref{HH}),
we note that $\mathcal{H}_y(k_y)$ determines gap closing on the edge because $\mathcal{H}_x(k_x)$ has the nonzero $w_x$.
$\mathcal{H}_x^{\mathrm{OBC}}$ has the topological zero-mode $\phi _x^{zero}$ now,
and $\mathcal{H}_y(k_y)$ has Bloch eigenstates $\psi _{n_y}(k_y)$ with the eigenvalues $E_{n_y}(k_y)$.
Therefore, an eigenvector of the ${H}^{\mathrm{EBC}}_x(k_y)$ can be obtained from $\phi _x^{zero} \otimes \psi _{n_y}(k_y)$
because it satisfies
\begin{align}
	{H}^{\mathrm{EBC}}_x(k_y)[\phi _x^{zero} \otimes \psi _{n_y}(k_y)]
	= E_{n_y}(k_y)[\phi _x^{zero} \otimes \psi _{n_y}(k_y)].
\end{align}
$\mathcal{H}_y(k_y)$ can effectively describe the edge states near zero energy
thanks to the nontrivial topology of $\mathcal{H}_x(k_x)$.
Hence, the gap closes on the edge when $\mathcal{H}_y(k_y)$ takes zero-valued eigenvalues.
%because $E_{y}^n(k_y)$ and $-E_{y}^n(k_y)$ appear in pairs thanks to the chiral symmetry.

We also analyze a topological phase transition from the phase with $(w_x, w_y)=(0,1)$ to the phase with $(1,1)$.
In this case, we show that gap closing happens on the edge normal to the $y$ direction.
The Hamiltonian with the edge boundary is 
\begin{align}
	{H}^{\mathrm{EBC}}_y(k_x)=\mathcal{H}_x(k_x) \otimes \Pi _y^{\mathrm{OBC}}+ 1_x \otimes \mathcal{H}_y^{\mathrm{OBC}}.
	\end{align}
Similarly, we can obtain an eigenstate of ${H}^{\mathrm{EBC}}_y(k_x)$ from $\psi _{n_x}(k_x) \otimes \phi _y^{zero}$
due to the nonzero $w_y$.
$\psi _{n_x}(k_x)$ is a Bloch function of $\mathcal{H}_x(k_x)$.
We can choose $\phi _y^{zero}$ to satisfy $\Pi _y^{\mathrm{OBC}}\phi _y^{zero}=\pm \phi _y^{zero}$
because this state is a zero-energy mode of $\mathcal{H}_y^{\mathrm{OBC}}$.
Therefore, the gap closes on the edge through the eigenvalues of $\mathcal{H}_x(k_x)$.
%As a result, the change of the topological invariant $\nu_{2D}$ is correspondent to the gap closings on the bulk and the edges.

\subsection{Construction of 3D SOTSMs and SOTIs}
Hereafter, we generalize the theory about the 2D SOTI to 3D systems with chiral symmetry
by adding translation symmetry in the $z$ direction.
Since the wave vector $k_z$ is added to Eq.~(\ref{kH}) in the 3D systems,
the bulk Hamiltonian is given by
\begin{align}
	H(\bm{k})=\mathcal{H}_x(k_x, k_z)\otimes \Pi _y+ 1_x \otimes \mathcal{H}_y(k_y, k_z). \label{TSM}
	\end{align}
Thus, we can regard $k_z$ as a new parameter leading to the topological phase transition
characterized by $\nu _{2D}(k_z)$.
The topological phase transition can occur in momentum space,
in which gap-closing points can emerge due to the chiral symmetry for the 3D system.
Therefore, the phase has the topological gap-closing points and hinge states at zero energy.
As a result, we can realize a 3D second-order topological phase by stacking the 2D SOTIs.

\begin{figure}[t]
	\includegraphics[width=8.5cm]{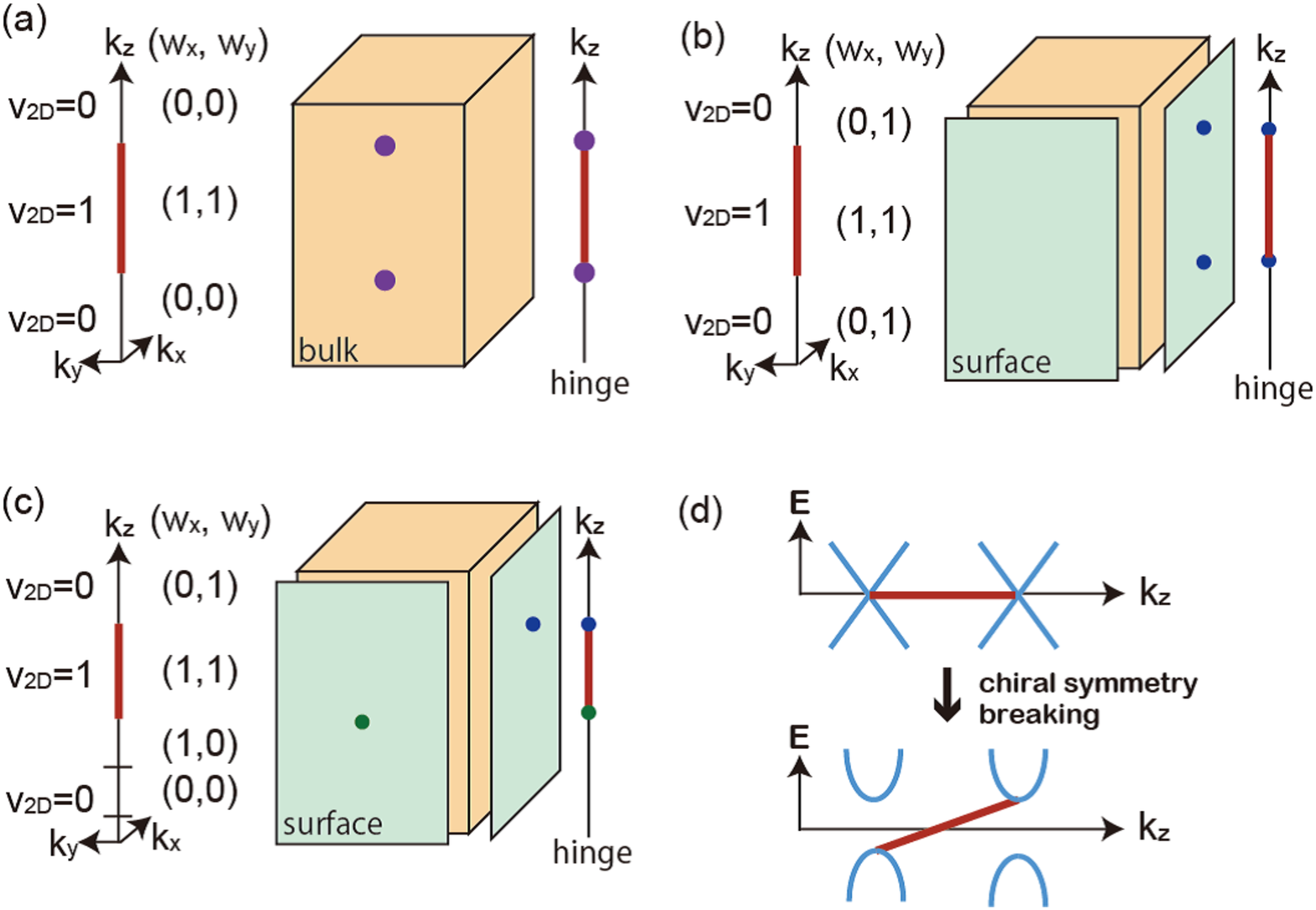}
	\caption{\label{tsm} (a)-(c) Schematic drawings of gapless points in the 3D second-order topological phases. 
		The dots are the gapless points. %, and the color difference represents various positions of the gap closing.
		The red lines indicate the topological hinge states.
		(a) The bulk gap closings give rise to the gapless points.
		(b) and (c) The gapless points appear from the surface gap-closings while the bulk is insulating.
		The surface gap closes when $\nu _{2D}(k_z)$ changes.
		(d) Schematic drawing of hinge band evolution for the topological phase transition to the 3D SOTI induced by broken chiral symmetry.
	}
\end{figure}

The positions of the gapless points depend on
how $\mathcal{H}_x(k_x, k_z)$ and $\mathcal{H}_y(k_y, k_z)$ give rise to the topological phase transitions.
Because $\nu _{2D}(k_z)$ is periodic for $k_z$, point nodes always appear in pairs.
If $\mathcal{H}_x$ and $\mathcal{H}_y$ change the winding numbers $w_x$ and $w_y$ simultaneously,
a SOTSM phase emerges in the bulk, as shown in Fig.~\ref{tsm}(a). 
The system shows zero-energy hinge states between the nodes projected onto the hinges
because of the nonzero $\nu _{2D}(k_z)$.
For example, the gapless node appears when $(w_x, w_y)$ changes from $(0,0)$ to $(1,1)$ in the 3D momentum space.

By contrast, the system can have gapless points on the surfaces.
While the bulk is insulating,
the surface gapless points coexist with topological hinge states.
There are two types of topological phase transitions by the change in $k_z$,
which results in analogs of surface topological semimetals.
We consider the first type where gapless points appear on the same surface.
In this type, either $w_x$ or $w_y$ changes in momentum space
while the other one is fixed to a nonzero value, for instance [Fig.~\ref{tsm}(b)].

In the second type, we can obtain gapless points on the different surfaces.
The point nodes can be found when $w_x$ and $w_y$ change in momentum space.
We note that this surface gapless structure is unique to the 3D systems
because 2D bulk topological semimetals necessarily have topological point nodes
in pairs in momentum space \cite{Chiu16, Matsuura13, Zhao16, Ahn17, Park17}. 
As an example, each surface can have a single gapless point
if $(w_x, w_y)$ changes as $(0,1) \rightarrow (0,0) \rightarrow (1,0) \rightarrow (1,1) \rightarrow (0,1)$,
as illustrated in Fig.~\ref{tsm}(c).
In both types, topological hinge states appear at the zero energy between the gapless points.

Next, we break chiral symmetry in the 3D SOTI with the gapless points on the surfaces.
The gap opens at the point nodes because the symmetry protection is absent.
Hence, we can discuss a surface Chern insulator due to massive Dirac cones \cite{Sitte12, Zhang13, Chen19, Haldane88}.
In the specific case, we can easily diagnose the existence of the gapless hinge states (see Appendix \ref{chiral}).
Typically, if the pairs become gapped by broken chiral symmetry,
they can contribute to the surface Chern numbers. % depending on the perturbation.
The chiral hinge states appear between the gapped surfaces on the trivial bulk.
Consequently, the system can show the 3D SOTI phase with chiral symmetry breaking [Fig.~\ref{tsm}(d)].

\section{Model} \label{model}
We study a lattice model to demonstrate the second-order topological phases realized by chiral symmetry,
and confirm our theory in the previous section.
The 2D lattice model is constructed from the Su-Schrieffer-Heeger (SSH) model \cite{Su79},
which can be transformed to the Benalcazar-Bernevig-Hughes model \cite{Benalcazar17S, Benalcazar17B}.

\subsection{The 2D SOTI model}
\begin{figure}[t]
	\includegraphics[width=8.5cm]{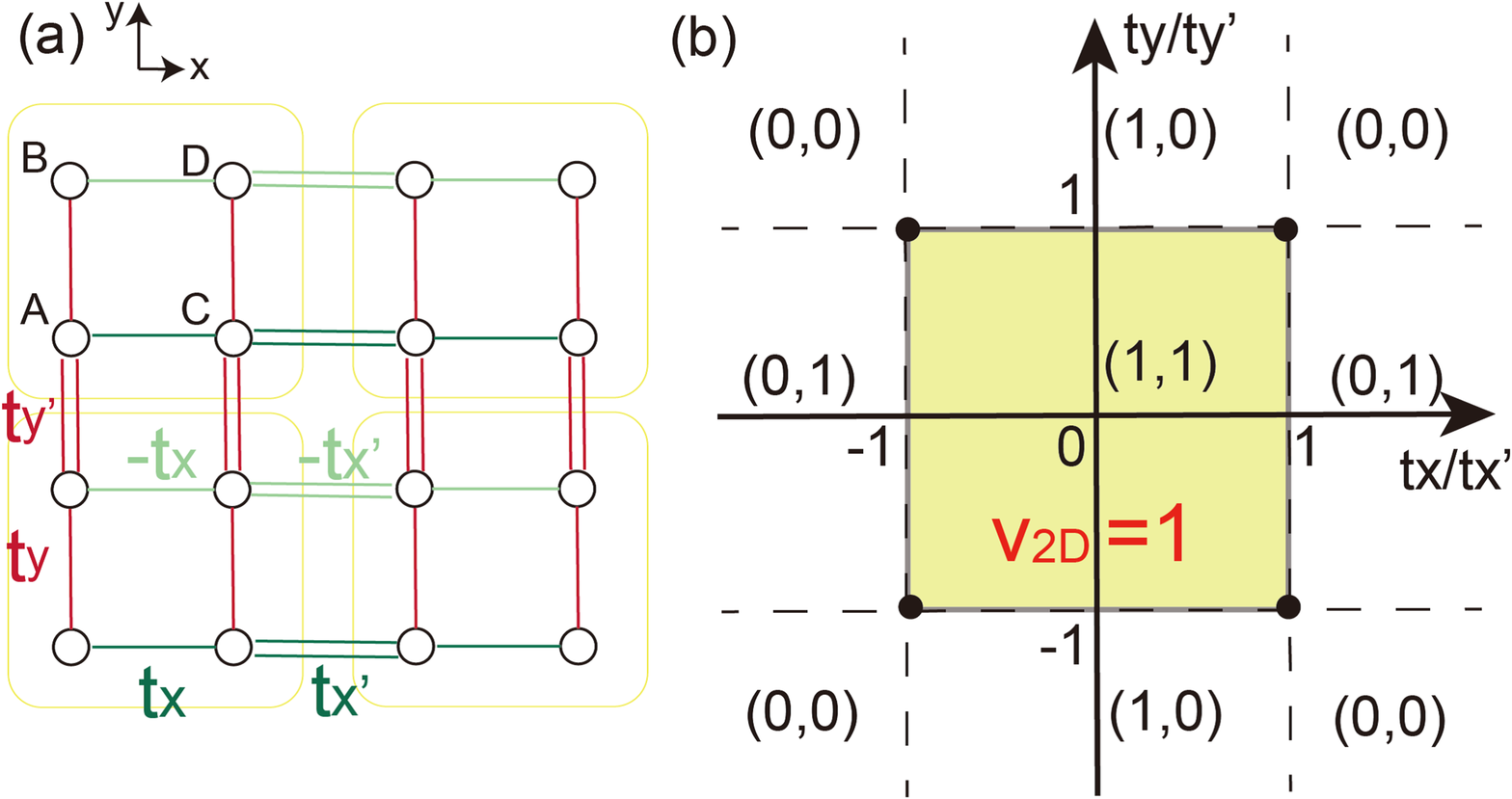}
	\caption{\label{phase} (a) Schematic drawing of the 2D model for the 2D SOTI.
		The unit cell consists of four sublattices.
		(b) Phase diagram of the 2D model. The yellow shaded region represents the topological phase.
		When $(w_x, w_y)$ gives the nonzero $\nu _{2D}$, the gap closing occurs. 
		The dots on the phase boundary represent the bulk gap closings.
	}
\end{figure}

We construct second-order topological phases and the Hamiltonian using the general method from Eq.~(\ref{kH}).
We study a 2D tight-binding model described as 
\begin{align}
	H&=\sum _{\bm{R}}[t_x(c^{\dagger}_{\bm{R}C}c_{\bm{R}A}-c^{\dagger}_{\bm{R}D}c_{\bm{R}B}+\mathrm{H.c.}) \notag \\
	&+t_y(c^{\dagger}_{\bm{R}B}c_{\bm{R}A}+c^{\dagger}_{\bm{R}D}c_{\bm{R}C}+\mathrm{H.c.}) \notag \\	
	&+t'_x(c^{\dagger}_{\bm{R}+\hat{x}A}c_{\bm{R}C}-c^{\dagger}_{\bm{R}+\hat{x}B}c_{\bm{R}D}+\mathrm{H.c.}) \notag \\
	&+t'_y(c^{\dagger}_{\bm{R}+\hat{y}A}c_{\bm{R}B}+c^{\dagger}_{\bm{R}+\hat{y}C}c_{\bm{R}D}+\mathrm{H.c.})],  \label{2DSSHreal}
\end{align}
where $\hat{x}$ and $\hat{y}$ are the unit vectors in the $x$ and $y$ directions, respectively.
$t_x,t_x',t_y$ and $t'_y$ are real hopping parameters [see Fig.~\ref{phase}(a)].
We set the lattice constants to unity.
The Hamiltonian in momentum space is
\begin{align}
	H(\bm{k})&=[(t_x+t'_x\cos k_x)\tau _x + t'_x\sin k_x\tau _y] \otimes \sigma _z \notag \\
	&+\tau _0\otimes [(t_y+t'_y\cos k_y)\sigma _x + t'_y\sin k_y \sigma _y], \label{2DSSH}
\end{align}
where $\sigma_{x,y,z}$ and $\tau _{x,y,z}$ are Pauli matrices acting on the sublattices,
and $\sigma _0$ and $\tau_0$ are the identity matrices.
The energy eigenvalues are 
\begin{align}
	E(\bm{k})=\pm \sqrt{\sum _{i=x,y}[(t_i+t'_i\cos k_i)^2+(t'_i\sin k_i)^2]}. \label{2DSSHenergy}
\end{align}
This model actually consists of the two SSH models represented by
\begin{align}
&\mathcal{H}_{j=x,y}(k_j) = \begin{pmatrix}
	0 & t_j+t'_je^{-ik_j}\\
	t_j+t'_je^{ik_j} & 0
\end{pmatrix}, 
\end{align}
and the chiral symmetry is given by
\begin{align}
&\Pi _{x,y} = \begin{pmatrix}
	1 & 0 \\
	0 & -1
\end{pmatrix}.
	\end{align}
Thus, this 2D model has chiral symmetry $\Pi = \tau _z \otimes \sigma _z$.

The topological invariant $\nu _{2D}$ can be calculated from the winding numbers $w_j$ for $\mathcal{H}_j(k_j)$.
They are given by
\begin{align}
	w_j=
	\begin{cases}
		1 & |t_j/t_j'|<1 \\
		0 & |t_j/t_j'|>1
		\end{cases}.
	\end{align}
Hence, we can obtain the phase diagram for the 2D model from Eq.~(\ref{inv}) [Fig.~\ref{phase} (b)].
When the corner boundary condition is imposed and $\nu _{2D}=1$,
the zero mode appears.
In the Appendix \ref{exactzero}, we show the analytic eigenvector localized at the corner in this model.

We note that our model in Eq.~(\ref{2DSSH}) is topologically equivalent to the Benalcazar-Bernevig-Hughes model 
which can have topological corner modes protected by two anticommuting mirror symmetries \cite{Benalcazar17S, Benalcazar17B}.
Alternatively, we can reinterpret the zero-energy corner modes
as a manifestation of the extrinsic second-order topology under the corner boundary condition in class AIII \cite{Hayashi19}.
Therefore, the corner modes can remain stable topologically even after the mirror symmetries are broken.
We investigate the stability by breaking the crystal symmetries in Appendix \ref{brc}.

\subsection{Stacked SOTI model}
We stack the 2D SOTI model to demonstrate 3D second-order topological phases.
While we preserve the chiral symmetry, we introduce terms depending on $k_z$.
The dependence on $k_z$ is determined by how the 2D SOTIs are stacked in the $z$ direction.

\begin{figure*}[t]
	\includegraphics[height=6.5cm]{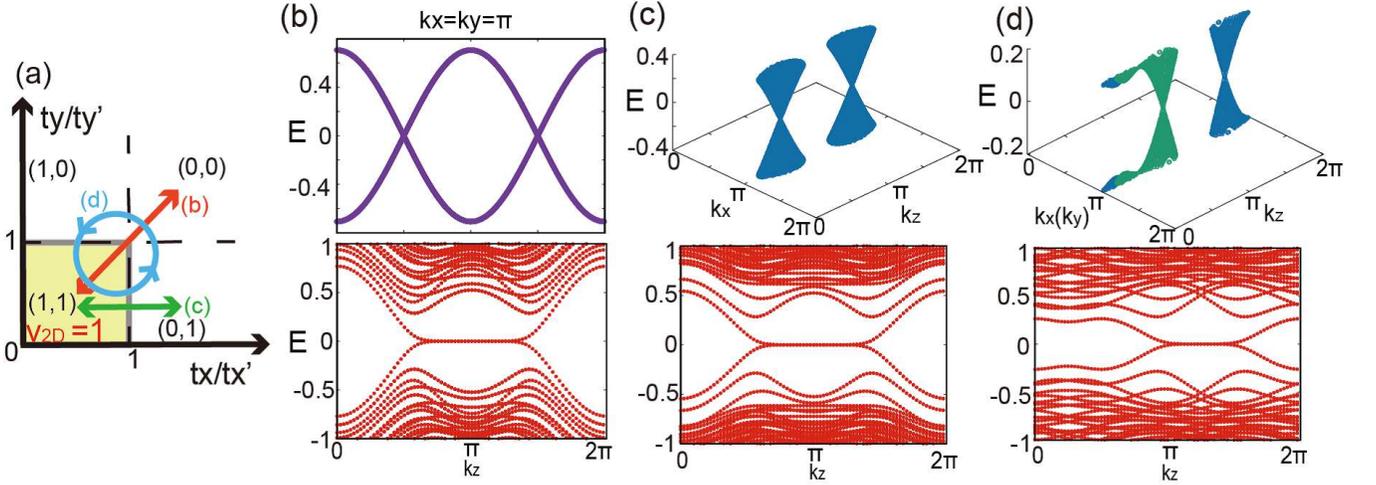}
	\caption{\label{bands}(a) Trajectories of the hopping parameters in the phase diagram.
		We put $t_x'=t_y'=1$ in the numerical calculations.
		The red (green) arrow indicates the case 
		where $t_x=t_y=1.0$ and $\chi _1=\chi _2=0.25$ for (b)
		[$t_x=1.0, t_y=0.4, \chi _1=0.25$, and $\chi _2=0$ for (c)].
		The blue circle represents the change in the hopping parameters
		with $t_x=t_y=0.5$, $\delta t_x=\delta t_y=0.3$, and $\eta =0.2$ for (d).
		(b)-(d) The top panels show gap-closing points from the topological phase transitions, 
		and the bottom panels show the hinge bands in a regular square column with the $15 \times 15$ unit cells.
		In (b), the bulk bands are calculated along the line $(k_x, k_y)=(\pi, \pi)$.
		The bulk has the point nodes at $k_z=\pi /2$ and $3\pi /2$ on the line. 
		The top panel in (c) shows bands for the surfaces normal to the $y$ direction.
		The surface bands have nodal points at $(k_x, k_z)=(\pi , \pi /2)$ and $(\pi , 3\pi /2)$.
		In (d), the blue and the green bands represent electronic structures
		for the surfaces normal to the $y$ and the $x$ directions, respectively.
		The gapless point in the blue (green) bands is located at $(k_x, k_z)=(\pi ,5\pi /3)~((k_y, k_z)=(\pi ,5\pi /6))$.
	}
\end{figure*}

First, we add two terms $2\chi _1\cos k_z\tau _x \otimes \sigma _z$ and $2\chi _2\cos k_z\tau _0 \otimes \sigma _x$
to the Hamiltonian in Eq.~(\ref{2DSSH}).
These terms effectively alter the hopping terms $t_x$ and $t_y$ to
$t_x+2\chi _1\cos k_z$ and $t_y+2\chi _2\cos k_z$, respectively.
Hence, the topological phase transition can happen in the momentum space.
When $t_x+2\chi _1\cos k_z=\pm t_x'~(|t_y/t_y'|\leq 1)$ and/or $t_y+2\chi _2\cos k_z=\pm t_y'~(|t_x/t_x'|\leq 1)$ can be satisfied,
the $k_z$ planes have gapless points.

Suppose that $t_x=t_y$, $t_x'=t_y'$ and $\chi _1 = \chi _2$.
In the parameter region, we can obtain a SOTSM with the bulk gapless points.
If the parameters change as the red arrow in Fig.~\ref{bands}(a),
two fourfold-degenerate points appear
at $(k_x, k_y, k_z)=(\pi , \pi , \pm \arccos \frac{t_x'-t_x}{2\chi _1})$ in the bulk.
The topological hinge states also exist between the projected point nodes, as shown in Fig.~\ref{bands}(b).

Next, we set $\chi _2=0$ to see a SOTI with gapless points on the surfaces.
Since $t_y$ is fixed, only $w_x$ can change in momentum space.
Therefore, we can realize gapless points on the surface perpendicular to the $y$ direction.
For example, we change $t_x$ as the green arrow in the diagram.
$w_x$ and $\nu _{2D}$ change when $t_x+2\chi _1\cos k_z=t_x'$.
Figure \ref{bands}(c) shows the two surface nodes at
$(k_x, k_z)=(\pi , \pm \arccos \frac{t_x'-t_x}{2\chi _1})$ and the zero-energy hinge states
between the two points projected to the hinges. 

Second, we add the new terms,
$(2\eta \cos k_z+\delta t_x)\tau _x \otimes \sigma _z$ and $(2\eta \sin k_z+\delta t_y)\tau _0 \otimes \sigma _x$,
instead of the previous two terms.
The hopping $t_x$ $(t_y)$ is effectively regarded as $t_x+\delta t_x +2\eta \cos k_z$ $(t_y+\delta t_y +2\eta \sin k_z)$.
If $t_x'=t_y'$, the modified hoppings form the circle in the diagram, as depicted in Fig.~\ref{bands}(a).
Thus, the gap closing can occur on the different surfaces
when the circle encloses a phase transition point accompanied by bulk gap closing.
We assume that the circle encloses the point $t_x/t_x'=t_y/t_y'=1$, and that $\eta /t_x' >0$.
Then, the point nodes appear
at $(k_x, k_z)=(\pi, -\arccos \frac{t_x'-t_x-\delta t_x}{2\eta})$ and $(k_y, k_z)=(\pi, \pi -\arcsin \frac{t_y'-t_y-\delta t_y}{2\eta})$
on the surfaces normal to the $y$ and the $x$ directions, respectively [Fig.~\ref{bands} (d)].
The anomalous distribution of the point nodes is allowed
because the gap closings both in the bulk and on the surfaces change the hinge topology,
similar to a mirror-symmetric topological quadrupolar semimetal \cite{Lin18}.
To annihilate the surface point nodes, the gap should close in the bulk.

We emphasize that our Hamiltonian does not need any crystal symmetries to realize the second-order topological phases.
The 2D chiral-symmetric model can also be characterized by the two mirror symmetries
for quantization of the nested Wilson loop \cite{Benalcazar17S, Benalcazar17B, Lin18}.
However, even if the mirror symmetries are broken,
the hinge flat-band states can survive as long as the chiral symmetry exists.
We can indeed see the protection of the chiral symmetry by breaking the mirror symmetries.
For example, 
we introduce the term 2$t_{w_1}\cos k_z \tau _y\otimes \sigma _z+2t_{w_2}\cos k_z\tau _0 \otimes \sigma _y$
for the case in Fig.~\ref{bands}(c).
If $t_{w_{1,2}} \ll t_{x,y}'$, the topological zero-modes are stable because $\nu _{2D}$ does not change [Fig.~\ref{TRB} (a)].

\subsection{SOTIs with broken chiral symmetry}
\begin{figure}[t]
	\includegraphics[height=3.7cm]{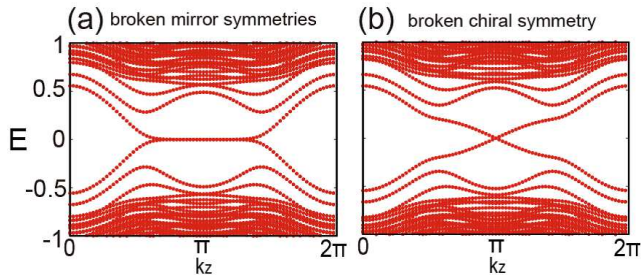}
	\caption{\label{TRB} The hinge bands for the case of Fig.~\ref{bands} (c) with broken symmetries.
		(a) The zero-energy hinge states survive without the mirror symmetries after the perturbation with $t_{w_1}=t_{w_2}=0.1$ is added.
		(b) The gapless chiral hinge states emerge by the perturbation with $t_m=0.1$. 
	}
\end{figure}

Finally, we break chiral symmetry to produce a 3D SOTI from the 3D SOTI with the surface gapless points.
The system belongs to class A.
By using our model, we here show that the SOTI can be realized easily from the surface gapless points in Fig.~\ref{bands} (c).
We add the perturbation $2t_m \sin k_z \tau _z \otimes \sigma _z$ in order to obtain the SOTI without chiral symmetry.
Figure \ref{TRB}~(b) shows the gapless hinge states which originate from the zero-energy states.

To understand the topological phase transition,
we investigate the surface normal to the $y$ direction from the discussion in Sec.\ref{SOTP} B.
Because $\mathcal{H}_y$ is topologically nontrivial,
%the surface states considered are represented effectively 
we can obtain the surface effective Hamiltonian near the zero energy by expanding
$\mathcal{H}_{xz}=(t_x+t_x'\cos k_x+2\chi _1\cos k_z)\tau _x+t_x'\sin k_x\tau _y + 2t_m\sin k_z\tau _z$
%$(t_x+t_x'\cos k_x+2\chi _1\cos k_z)\tau _x+t_x'\sin k_x\tau _y + 2t_m\sin k_z\tau _z$
(see also Appendix \ref{chiral}).
The effective Hamiltonian describes the surface Dirac points with the mass induced by the broken chiral symmetry.

%We can calculate the surface Chern number from $\mathcal{H}_{\mathrm{eff}}$. 
We can calculate the surface Chern number from the surface Hamiltonian.
The effective Hamiltonian near the gapless points $\bm{k}^{\pm}\equiv (\pi , \pm \arccos \frac{t_x'-t_x}{2\chi _1})$ is given by
\begin{align}
	\mathcal{H}_{\mathrm{eff}}^{\pm}(\bm{q})=-(2\chi _1\sin k_z^{\pm})q_z\tau _x-t_x'q_x\tau _y+2t_m\sin k_z^{\pm}\tau _z,
\end{align}
where $\bm{q}=(q_x, q_z)$ is the wave vector measured from $\bm{k}^{\pm}$.
By the perturbation term, the Dirac cones at $\bm{k}^{\pm}$ obtain the mass gap described as $2t_m\sin k^{\pm}_z\tau _z$.
Because $\sin k_z^+>0$ and $\chi _1/t_x'>0$ in the calculation,
the surface has a Chern number of $-\mathrm{sgn}(t_m)$.
As a result, the system becomes the SOTI phase,
which is consistent with the discussion in Appendix \ref{chiral}.

\section{Conclusion and discussion} \label{con}
In the present paper,
we have shown that various chiral-symmetric second-order topological insulators and semimetal are realizable.
We have suggested the method to construct the second-order topological phases protected by chiral symmetry.
The theory reveals that there are several types of chiral-symmetric second-order topological phases in three-dimensional systems,
and that the topological phases can be obtained from stacked two-dimensional second-order topological insulators with chiral symmetry.
The zero-energy hinge states are stable topologically by chiral symmetry.
Particularly, we have discovered the second-order topological insulator,
which has not only hinge bands but also a single gapless point on the surface thanks to chiral symmetry.
In general, this surface gapless point is not allowed in the two-dimensional bulk
because both the bulk and surfaces need to contribute topologically.
We have also shown that broken chiral symmetry can yield the second-order topological insulator
from the surface gapless points in the chiral-symmetric second-order topological insulators.
Our method tells us how to give extrinsic second-order topology easily in classes A and AIII.

Importantly, if any Hamiltonian with chiral symmetry can be continuously deformed to Eq.~(\ref{kH}) or (\ref{TSM})
without the bulk and surface gap closings,
we can distinguish whether the system has nontrivial second-order topology
because the band topology is unchanged \cite{Hayashi18, Hayashi19}.
Namely, our theory is useful even though the Hamiltonian does not take the form in Eq.~(\ref{kH}) or (\ref{TSM}).
%the gapless hinge states are stable topologically unless the band gap closes.
%Therefore, our proposed topological phases universally appear.
Thus, our model is available for photonic and phononic crystals and electric circuits
\cite{Imhof18, Serra18, Peterson18, Ni19, Fan19} because their structures are flexibly controllable.
Theoretically, corner modes in class BDI are suggested in photonic systems
composed of waveguides and of coupled ring resonators with negative hopping amplitudes \cite{Li18}.
We also expect that our second-order topological phases can be constructed in waveguide arrays,
and that the corner modes are detectable
through measurement of the outgoing intensity distribution of light injected into a waveguide.
Moreover,
electric circuits have the possibility of experimental realization of the second-order topological phases protected by chiral symmetry,
because capacitors and inductors can provide band structures of the circuit Laplacian \cite{Imhof18, Lee18, Helbig19}. 
%The capacitances and inductivities corresponds to hopping parameters.
The nontrivial second-order topological modes in the electric circuits are measurable by the impedance response
since the zero modes contribute to large resonances at the boundaries such as corners.
Therefore, our model is feasible in systems with the easily controllable structures.

Additionally, topological stability of gapless nodes is characterized locally
by the change of a topological invariant in momentum space \cite{Chiu16, Matsuura13, Zhao16}.
Therefore, the nodal structures in our paper are allowed in other classes with chiral symmetry.
Actually, second-order topological phases have been predicted in superconductors
\cite{Langbehn17, Geier18, Luka19, Khalaf18B, Shapourian18, Zhu18, Yan18, Wang18B, Wang18L, Hsu18, Liu18, Bultinck19, Wu19, Zhang19, Peng19, Ghorashi19}.
Some of the topological superconductors are understandable in view of extrinsic second-order topology \cite{Peng19, Ghorashi19}.
Recently, robust Majorana hinge states were studied in a second-order topological superconductor
as long as the particle-hole symmetry is preserved \cite{Ghorashi19}.
%A way to realize an extrinsic second-order superconductor has been also theoretically suggested
%in an antiferromagnet topological insulator with proximity superconducting gap \cite{Peng19}.
Thus, the extrinsic second-order topological nodal phases protected by chiral symmetry are expected to appear
in superconductors with time-reversal symmetry.

\begin{acknowledgments}
We are grateful to R. Takahashi and S. Murakami for helpful discussion.
This work is supported by World Premier International Research Center Initiative (WPI), MEXT, Japan, JPSJ Grant-in-Aid Scientific Research on Innovative Areas “Discrete Geometric Analysis for Materials Design” (Grant No. 17H06460 and No. 17H06461),
and JSPS KAKENHI Grants No. JP19K14545 and No. 16K05412, and JST-CREST (JPMJCR18T1), and JST-PRESTO (JPMJPR19L7). 
\end{acknowledgments}

\appendix
\section{Expression of the winding number}\label{winding}
To describe the topological invariant $\nu _{2D}$ in a self-contained way,
we explicitly write the winding number $w$
for the 1D Hamiltonian with chiral symmetry.
To do so, we introduce the following $Q$ matrix:
\begin{align}
	Q(k)=1-2\sum_{E_n<0}\ket{\psi _n(k)}\bra{\psi _n(k)},
	\end{align}
where the sum is taken over the eigenstates $\ket{\psi _n(k)}$ below zero energy.
In the basis where the chiral symmetry is diagonal, 
the $Q$ matrix can be rewritten as
\begin{align}
	Q(k)=
	\begin{pmatrix}
		0 & q(k) \\
		q^{\dagger}(k) & 0
	\end{pmatrix}.
	\end{align}
By using the off-diagonal component $q(k)$, the winding number is defined as \cite{Schnyder08, Ryu10, Chiu16}
\begin{align}
	w=\frac{i}{2\pi}\int _{0}^{2\pi}dk\mathrm{Tr}[q^{-1}\partial _kq].
	\end{align}
The winding number $w$ corresponds to the number of topological edge states in the 1D chiral-symmetric system.

\section{Model for chiral hinge states}\label{chiral}
In this appendix, we give a 3D Hamiltonian which shows chiral hinge states without the protection of any symmetries.
The Hamiltonian is beneficial to obtain the 3D extrinsic SOTI in class A.
We consider the following bulk Hamiltonian \cite{Hayashi18}:
\begin{align}
	H_A(\bm{k})=\mathcal{H}_{xz}(k_x,k_z) \otimes \Pi _y + 1_x \otimes \mathcal{H}_y(k_y). \label{bulkA}
	\end{align}
Here, $\mathcal{H}_{xz}(k_x,k_z)$ is a Hermitian matrix without symmetry,
and $1_x$ has the same size as $\mathcal{H}_{xz}(k_x,k_z)$.
$\mathcal{H}_y(k_y)$ is the chiral-symmetric Hamiltonian explained in Sec.~\ref{SOTP}.
We assume that the bulk and the surface gaps are open at zero energy.
Because Eqs.~(\ref{bulkA}) and (\ref{kH}) have the same structure,
we discuss the topological hinge states in a manner similar to Sec.~\ref{SOTP}
by regarding $\mathcal{H}_{xz}(k_x,k_z)$ as a 2D bulk Hamiltonian.
Therefore, we define a Chern number $\mathrm{Ch} _{xz}$ for $\mathcal{H}_{xz}(k_x,k_z)$,
which gives the 1D chiral edge states $\psi ^{chiral}$.
The Chern number $\mathrm{Ch} _{xz}$ is defined for the bulk bands below zero energy
since the origin of the energy can be shifted.
 
To see chiral hinge states in the 3D SOTI, we investigate the semi-infinite Hamiltonian with one hinge:
\begin{align}
	H_A^{\mathrm{HBC}}(k_z)
	=\mathcal{H}_{xz}^{\mathrm{OBC}}(k_z) \otimes \Pi _y^{\mathrm{OBC}} + 1_x^{\mathrm{OBC}} \otimes \mathcal{H}_y^{\mathrm{OBC}}. 
	\end{align}
We maintain the translation symmetry in the $z$ direction.
In the same way as in Sec.~\ref{SOTP}, 
we can obtain topological hinge states $\psi ^{chiral}(k_z) \otimes \phi _y^{zero}$ with the gapless dispersion
if $\mathrm{Ch} _{xz}$ and $w_y$ are nonzero.
Thus, we can introduce a $\mathbb{Z}$ topological invariant $I=\mathrm{Ch}_{xz}w_y$,
which characterizes the number of gapless hinge states \cite{Hayashi18}.
Realistically, $\mathcal{H}_y$ can depend on $k_z$.
However, since $k_z$ can be interpreted as a parameter in $\mathcal{H}_y(k_y)$,
gapless hinge states can be detected if the model can be adiabatically deformed to Eq.~(\ref{bulkA}) .

In comparison with our model in Sec.~\ref{model} C,
%$\mathcal{H}_{xz}$ corresponds to $\mathcal{H}_{\mathrm{eff}}$.
$\mathcal{H}_{xz}$ is given by $\mathcal{H}_{xz}=(t_x+t_x'\cos k_x+2\chi _1\cos k_z)\tau _x+t_x'\sin k_x\tau _y + 2t_m\sin k_z\tau _z$.
The massive Dirac cones on the surface give the nonzero Chern number $\mathrm{Ch}_{xz}=-\mathrm{sgn}(t_m)$.
As a result, we obtain the chiral hinge states from the SOTI with the broken chiral symmetry.

On the other hand, the phase transition from a SOTSM to a SOTI with chiral hinge states was also suggested
in 3D systems with parity symmetry \cite{Wang18, Ahn18}.
The SOTSM hosts nodal lines topologically protected by parity-time symmetry in pairs,
giving nearly flat hinge bands between the nodal lines.
Time-reversal symmetry breaking in the SOTSM realizes the 3D SOTI,
and the nearly flat hinge states become the chiral hinge states in the SOTI.
The chiral hinge states are intrinsically $\mathbb{Z}_2$ protected by the parity symmetry \cite{Wang18, Ahn18, Geier18, Khalaf18X, Khalaf18B}.
By contrast, our topological phase transition stems from broken chiral symmetry,
and the induced chiral hinge states are characterized by the extrinsic second-order topology.

\section{Zero-energy states of the models}\label{exactzero}
%We investigate the zero-energy eigenvector of the chiral-symmetric second-order topological phases
%to see how the corner and the hinge modes are localized.
For the model in Eq.~(\ref{2DSSHreal}),
we can analytically compute the topological zero-energy eigenstate.
Thus, we can see how the zero-energy modes are localized.
The eigenstate can be obtained from zero modes of the SSH model.
Thus, we focus on the SSH model corresponding to $\mathcal{H}_{j=x,y}$ in Eq.~(\ref{2DSSH}),
which is topologically nontrivial if $|t_j/t'_j|<1$.
Let us label positions of the unit cell as $(n_x, n_y)~(n_{x,y}\geq 0)$ in the 2D semi-infinite real space with one corner
[see also Fig.~\ref{phase}(a)].
Since we can separately calculate the two SSH models in our model,
we denote components of the eigenvector to represent the $n_j$-th unit cell with the two sublattices
as $\bm{\phi} _{n_j} = (\phi _{n_j,1}, \phi _{n_j,2})^t~(n_j\geq 0)$.
The boundary condition is given by $\phi _{-1,2}=0$.

The components of the zero-mode satisfy
\begin{align}
	&t'_j\phi ^{zero}_{(n_j-1),2}+t_j\phi ^{zero}_{n_j,2}=0, \notag \\
	&t_j\phi ^{zero}_{n_j,1}+t'_j\phi ^{zero}_{(n_j+1),1}=0. \label{recurrence}
\end{align}
We can find the solution which decays as $n_j$ becomes larger when $|t_j/t'_j|<1$.
From Eqs.~(\ref{recurrence}), the solution is 
\begin{align}
	\bm{\phi} ^{zero}_{n_j} \propto \biggl(-\frac{t_j}{t'_j}\biggr) ^{n_j}
	\begin{pmatrix}
		1 \\
		0
	\end{pmatrix}.
\end{align}
Thus, the zero-energy state $\bm{\phi} ^{zero}$ for Eq.~(\ref{2DSSHreal}) can be obtained from
\begin{align}
	\bm{\phi} ^{zero} _{n_x, n_y} 
	\propto \biggl(-\frac{t_x}{t'_x}\biggr) ^{n_x}\biggl(-\frac{t_y}{t'_y}\biggr) ^{n_y}
	\begin{pmatrix}
		1 \\
		0 
	\end{pmatrix}\otimes
	\begin{pmatrix}
		1 \\
		0
	\end{pmatrix}.	\label{corzero}
\end{align}
Hence, the zero-energy mode is exponentially localized at the corner.

We can also describe zero-energy states of the 3D second-order topological phases in Sec.~\ref{model} C.
We give the hopping parameters $k_z$ dependence, and write them as $t_j+f_j(k_z)$.
The 3D model hosts the topological flat band if $|[t_x+f_x(k_z)]/t'_x|<1$ and $|[t_y+f_y(k_z)]/t'_y|<1$.
Since the system is periodic in the $z$ direction, 
we can consider the zero modes by fixing $k_z$ and replacing $t_j$ with $t_j+f_j(k_z)$ in Eq.~(\ref{corzero}).
As a result, we can see that the zero modes are localized at the hinge.
When $|[t_x+f_x(k_z)]/t'_x|=1$ and/or $|[t_y+f_y(k_z)]/t'_y|=1$,
the hinge flat band merges into the correspondent gapless points.

\section{Topological stability of the corner modes without protection of crystal symmetries}\label{brc}
Our 2D SOTI Hamiltonian in Eq.~(\ref{HH}) does not need protection of crystal symmetries thanks to the chiral symmetry.
Here, we show the topological stability of the corner modes by breaking crystal symmetries in the model in Eq.~(\ref{2DSSHreal}).
This also describes the robustness of hinge states in the 3D chiral-symmetric second-order nontrivial phases,
although the translation symmetry is necessary for the $k_z$ dependence.
Because the hinge states are characterized locally in momentum space by $\nu_{2D}(k_z)$,
we can understand them as corner modes in the 2D chiral-symmetric SOTI parametrized by $k_z$.
Therefore, we focus on the corner modes in the 2D SOTIs.

\begin{figure}[h]
	\includegraphics[width=8.5cm]{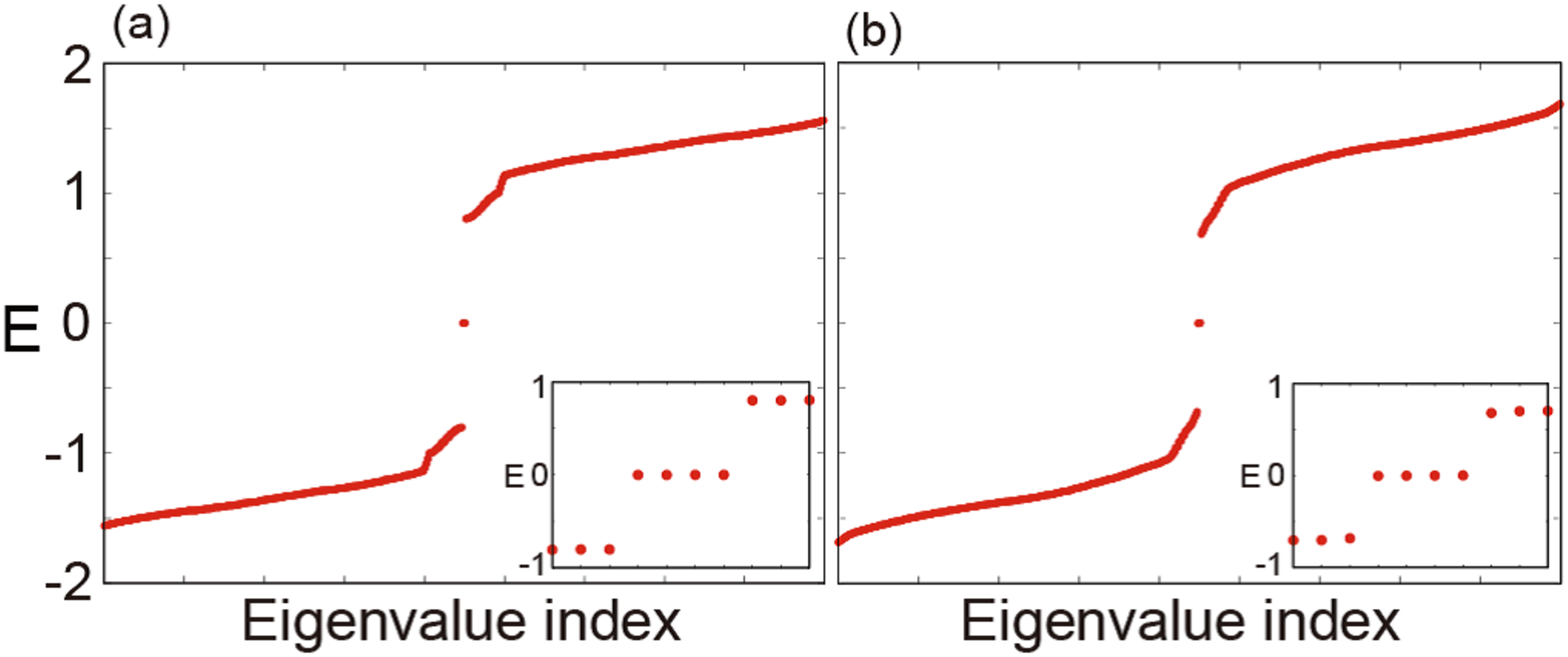}
	\caption{\label{cornerm} Energy eigenvalues of the 2D SOTI model in a regular square with the $15 \times 15$ unit cells.
	The insets show zero-energy eigenvalues of the corner modes. We put $t_x'=0.9, t_y'=1, t_x=0.1,$ and $t_y=0.2$ for the calculations.
	(a) shows the corner modes without the perturbation.
	(b) shows the topological stability of the corner modes in the presence of the perturbation with $t_{p_x}=0.2, t_{p_y}=0.1,$ and $t_p=0.05$.
	}
\end{figure}

The chiral-symmetric model has twofold-rotational symmetry, and two mirror symmetries with respect to the $x$ and the $y$ axis
if $t_x\neq t_y$ and $t_x'\neq t_y'$ in addition to time-reversal and particle hole symmetries.
To break the symmetries except the chiral symmetry, we add a perturbation term,
$H_p=\sum _{\bm{R}}\sum_{\alpha \beta}
[t_{p_x}(\tau _y \otimes \sigma _z)_{\alpha \beta}+t_{p_y}(\tau _0 \otimes \sigma _y)_{\alpha \beta}]
c^{\dagger}_{\bm{R}\alpha}c_{\bm{R}\beta}
+\sum _{\bm{R}}t_p[c^{\dagger}_{\bm{R}+\hat{x}B}c_{\bm{R}A}+c^{\dagger}_{\bm{R}+\hat{x}C}c_{\bm{R}D}+\mathrm{H.c.}]$ in Eq.~(\ref{2DSSHreal}),
where the summation of $\alpha$ and $\beta$ runs over sublattices $A$, $B$, $C$, and $D$.
The perturbation preserves the chiral symmetry.
Unless the additional term closes the bulk and edge gaps in the nontrivial phase,
the corner modes are robust even if the Hamiltonian does not have the form in Eq.~(\ref{kH}).
Figure \ref{cornerm} shows topological corner modes in the model.
As shown in Fig.~\ref{cornerm}(b), the corner modes survive after the perturbation breaking the crystal symmetries is added.
Hence, we can see the topological stability of the corner modes due to the chiral symmetry.

\bibliographystyle{apsrev4-1}% Produces the bibliography via BibTeX.
\bibliography{HOTP}

\begin{thebibliography}{78}
\bibitem [{com()}]{comment}%
  \BibitemOpen
  \href@noop {} {\bibinfo  {journal} {We especially use superscripts
  $\mathrm{CBC}, \mathrm{EBC}$,and $\mathrm{HBC}$ for the Hamiltonians when the
  corner, edge, and hinge boundary conditions are imposed, respectively}\
  }\BibitemShut {NoStop}%
\end{thebibliography}

\end{document}